\documentclass[journal]{IEEEtran}

\usepackage{amsmath,amssymb}
\usepackage{amsthm}
\usepackage{graphicx}
\usepackage{booktabs}
\usepackage{multirow}
\usepackage{array}
\usepackage{siunitx}
\usepackage{cite}
\usepackage{algorithm}
\usepackage{algorithmic}
\usepackage{url}
\usepackage{xcolor}
\usepackage{makecell}

\newcommand{\vct}[1]{\boldsymbol{#1}}
\newcommand{\vp}{\vct{p}}
\newcommand{\vd}{\vct{d}}
\newcommand{\vz}{\vct{z}}
\newcommand{\vu}{\vct{u}}
\newcommand{\vg}{\vct{g}} 
\newcommand{\vf}{\vct{f}} 
\newcommand{\vtheta}{\vct{\theta}} 
\newcommand{\vLS}{\vct{\mathsf{sh}}}

\newcommand{\mtx}[1]{\boldsymbol{#1}}
\newcommand{\mA}{\mtx{A}}
\newcommand{\mB}{\mtx{B}}

\newcommand{\mE}{\mtx{E}}

\newcommand{\mPhi}{\mtx{\Phi}}
\newcommand{\mGamma}{\mtx{\Gamma}}

\newcommand{\Ind}[1]{\mathbf{1}\left[#1\right]}

\newcommand{\EENS}{\mathsf{EENS}}
\newcommand{\ENS}{\mathsf{ENS}}

\newcommand{\ETC}{\mathsf{ETC}}
\newcommand{\VOLL}{\mathsf{VOLL}}
\newcommand{\SA}{\mathsf{SA}}
\newcommand{\ESA}{\mathsf{ESA}}

\title{Transmission Expansion with Large, Discrete\\ Data Center Realization Uncertainty}

\author{Lin~Dong,~\IEEEmembership{Graduate Student Member,~IEEE,}
        Samuel~Talkington,~\IEEEmembership{Member,~IEEE,}
        and~Le~Xie,~\IEEEmembership{Fellow,~IEEE}%
\thanks{Lin~Dong is with the Department
of Electrical and Computer Engineering, Texas A\&M University, College Station,
TX, 77840 USA (e-mail: \texttt{lin.dong@tamu.edu}).}%
\thanks{Samuel~Talkington is with the School of Engineering and Applied Sciences, Harvard University, Allston, MA 02134 USA, and is also with the Department of Electrical Engineering and Computer Science, University of Michigan, Ann Arbor, MI 48109 USA (e-mail: \texttt{talks@umich.edu}).}%
\thanks{Le~Xie is with the School of Engineering and Applied Sciences, Harvard University, Allston, MA 02134 USA (e-mail: \texttt{xie@seas.harvard.edu}).}%
}

\begin{document}

\maketitle

\begin{abstract}
Large data center projects introduce a distinct transmission planning uncertainty because of their large scale, discrete project realization nature, and strong locational impacts on the transmission network. On the one hand, conventional deterministic treatments can either overbuild the network by assuming that all candidate projects materialize; on the other, they can leave the system underprepared by prematurely screening out projects. This paper develops a project-realization-based framework for systematically modeling, planning, and evaluating transmission expansion under uncertain data center realization. Candidate data center projects are represented through alternative realization scenarios and associated probabilities, enabling different planning strategies to be modeled and quantitatively assessed under a common framework. Within this framework, numerical case studies demonstrate the effectiveness of the proposed framework and the benefits of stochastic planning approach. The results show that stochastic planning can use realization probabilities to hedge multiple plausible futures, providing a more balanced tradeoff between transmission investment and reliability exposure than deterministic approaches under correctly specified and misspecified realization probabilities. 

\end{abstract}


\section{Introduction}

\IEEEPARstart{L}{arge} loads, particularly large data centers, are emerging as an important source of long-term power system planning uncertainty. The North American Electric Reliability Corporation (NERC) characterizes large loads based on their potential impacts on bulk power system, while the Federal Energy Regulatory Commission (FERC) proposed a 50 MW threshold for certain transmission-connected loads \cite{NERC2025LargeLoads}, \cite{FERC2026LargeLoads}. Recent assessments project rapid growth in U.S. data center electricity demand, while also showing a wide range of possible future trajectories because the timing, scale, and completion of announced projects remain uncertain \cite{LBNL2024DC}, \cite{EPRI2026DC}, \cite{Chen2026AIGrid}. In addition to their aggregate energy impact, individual hyperscale facilities can reach hundreds of megawatts or more and may be geographically concentrated, creating location-specific transmission requirements that differ fundamentally from smooth aggregate load-growth uncertainty \cite{ESIG2026LargeLoad}. A proposed project may proceed as expected, be delayed or modified, or fail to materialize. Consequently, different combinations of realized data center projects can induce substantially different congestion patterns and transmission reinforcement needs.

This uncertainty creates an overbuilding--underbuilding dilemma for transmission planning. Treating every candidate project as certain can justify substantial infrastructure that may later be underutilized if some projects do not proceed. Conversely, excluding less-mature projects can leave the system underprepared when those projects eventually materialize. This concern is increasingly recognized in industry: large-load development timelines can be much shorter than transmission planning and construction timelines, while a significant portion of proposed projects may never materialize \cite{ESIG2026LargeLoad}. The planning challenge is therefore not only to forecast aggregate demand growth, but also to determine how much transmission infrastructure should be committed before the realization status of individual large-load projects is known.

In practice, utilities and system operators increasingly use project readiness and commitment information to distinguish credible large-load requests from speculative ones. For example, PSE\&G has assessed project-specific realization likelihoods using factors including site control, construction complexity and equipment availability \cite{PSEG2026Load}. ERCOT's Batch Zero process similarly applies study-maturity and commitment requirements to determine which large-load projects advance through system-wide transmission studies \cite{ERCOT2026BatchZero}. Such criteria-based screening improves planning inputs and is commonly translated into deterministic include-or-exclude assumptions, which does not eliminate project realization uncertainty. This approach is simple and practical, but it can lead to abrupt changes in the transmission plan when the actual realization differs. This motivates a project-level probabilistic treatment of data center realization uncertainty.

The literature on data center planning has increasingly considered the interaction between geographically distributed data centers and power system infrastructure. Existing studies have investigated data center siting and configuration, co-planning with energy storage, workload flexibility, and demand response \cite{Wang2023DCLocation}, \cite{Guo2021DCBESS}, \cite{Zeng2024DCPlanning}, \cite{kim2026flexibility}, \cite{kim2026theoretical}, \cite{He2025Shiftable}. More recently, flexible large load siting has been explicitly integrated into stochastic capacity-expansion planning with coordinated generation, storage, and transmission investment \cite{Valencia2027LargeLoad}. Data center flexibility has also been incorporated directly into generation and transmission expansion planning, where spatial and temporal load regulation can affect long-term expansion decisions \cite{Chen2021GTEP}. These studies establish that the location, operating flexibility, and supporting infrastructure of large data centers can materially affect power system planning decisions.

Uncertainty-aware formulations have further extended the literature. Vafamehr \emph{et al.} considered uncertainties in both electricity and data networks \cite{Vafamehr2019}. Distributionally robust and chance-constrained approaches have also been applied to integrated energy-system planning with data centers under wind power uncertainty \cite{Li2023DCDRO}. More recently, Shao \emph{et al.} developed a stochastic long-term joint planning framework for data centers and power systems that captures uncertainty in electricity demand and computing-workload growth with short-term operational and renewable variations \cite{Shao2025PJM}. These studies demonstrate the growing role of uncertainty-aware planning for data centers. In parallel, transmission expansion planning (TEP) under uncertainty is itself well-established \cite{Cao2025TEPReview}. Scenario-based stochastic TEP uses probability information to balance investment and operational outcomes across alternative futures \cite{Park2015STO,Zhan2017STO}, whereas robust TEP protects against adverse realizations within prescribed uncertainty sets and reduces dependence on precisely specified probability distributions \cite{Jabr2013Robust,Garcia2017DynamicRobust}. Risk-based and chance-constrained formulations provide additional mechanisms for controlling reliability exposure \cite{Qiu2016Risk,Li2018Chance}.

Despite these advances, the uncertainty represented in most existing data center planning studies differs from the project-realization problem considered here. Prior work primarily concerns where data center capacity should be located, how much capacity should be constructed, how workloads can be shifted, or how supporting generation, storage, and grid capacity should be co-optimized. However, a distinct planning problem arises when the transmission planner faces a known set of individually proposed, location-specific large-load projects whose requested locations and demand sizes are given, but whose eventual realization is uncertain. To the best of our knowledge, comparatively limited attention has been given to address this type of challenge. One related supply-side attempt is~\cite{yang2005chance}, which represents the realization of candidate generation projects using probabilities in a chance-constrained transmission expansion planning model. However, that study focuses on generation-side uncertainty and does not systematically analyze the impact of realization probability.

This paper addresses this challenge through a unified probabilistic modeling, planning, and evaluation framework that embeds large, discrete data center realization uncertainty. Candidate data centers are represented through binary realization states and assumed realization probabilities, allowing deterministic screening method, stochastic planning, and robust planning to be compared under a common realization structure. 
The main contributions are summarized as follows.

\begin{itemize}
    \item \textit{Project-level realization representation for large discrete data center uncertainty:}
    Candidate data center projects are represented by their locations, demand sizes, binary realization states, and realization probabilities, allowing deterministic planning, stochastic planning, and robust planning to be compared under a common realization structure. 

    \item \textit{Integrated planning framework under a common realization structure:}
    The same realization representation is embedded in multiple transmission planning methods, including deterministic planning, stochastic planning, and robust planning. This common structure enables fair and consistent comparison of different planning approaches.

    \item \textit{Unified ex-post performance assessment:}
    Expansion plans are compared using a common evaluation framework that quantifies investment, expected total cost, expected energy not served, and subsequent adaptation, defined as the corrective transmission expansion required after the actual data center realization becomes known. Cross-probability evaluation further measures how planning decisions degrade when assumed realization probabilities differ from the true probabilities.
\end{itemize}

The remainder of the paper is organized as follows. Section~\ref{sec:framework} introduces the proposed framework and the data center project realization model that forms its uncertainty basis. Sections~\ref{sec:planning} and~\ref{sec:evaluation} present the transmission planning formulations and ex-post evaluation framework, respectively. Section~\ref{sec:numerical} presents the numerical studies on two representative transmission systems. Section~\ref{sec:discussion} discusses the limitations, and Section~\ref{sec:conclusion} concludes the paper.

\section{Framework Overview and Data Center Realization Modeling}
\label{sec:framework}

\subsection{Framework Overview}

The uncertainty considered in this study differs from conventional load-forecast uncertainty in two fundamental aspects. First, candidate data centers represent large, non-marginal load additions relative to the existing demand and transfer capability of the areas to which they interconnect. The realization of a single project can therefore materially alter local power balances, transmission congestion patterns, and expansion requirements. Second, this uncertainty is inherently \emph{discrete} at the project realization level: a candidate project either materializes or does not, causing an entire block of demand to appear or disappear at a specific network location. Consequently, future demand uncertainty is represented not as a continuous perturbation of an aggregate load forecast, but as a finite set of spatially distinct combinations of large project realizations.

\begin{figure}
    \centering
    \includegraphics[width=1.0\linewidth]{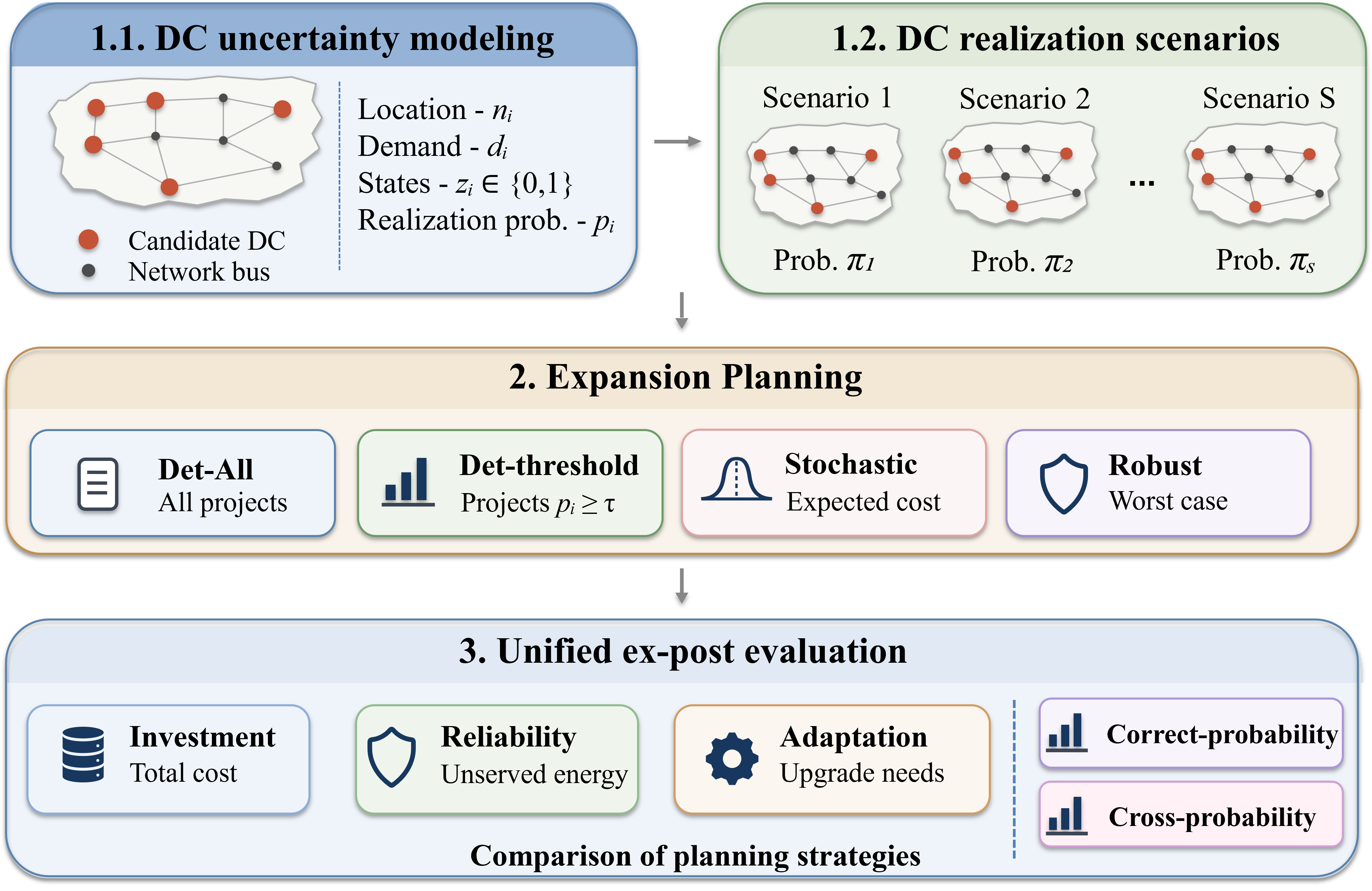}
    \caption{Overall modeling, planning, and evaluation framework for uncertain data center realization.}
    \label{fig:framework}
\end{figure}

To represent this uncertainty explicitly, this paper develops a project-realization-based framework that separates the overall planning process into three major interconnected components as illustrated in Fig.~\ref{fig:framework}: (i) data center realization modeling, which represents uncertain candidate projects through their locations, demand sizes, binary realization states, and assumed realization probabilities; (ii) probability-aware transmission planning, in which alternative planning paradigms use the same underlying realization information in different ways; and (iii) unified ex-post evaluation, in which the resulting expansion plans are assessed under a common realization environment.

The framework first converts project-level information into a structured set of realization scenarios. These scenarios provide a common uncertainty basis for alternative transmission planning strategies, while the resulting expansion plans are subsequently evaluated under the same ex-post realization environment. This structure separates uncertainty representation, planning decisions, and performance evaluation, enabling consistent comparison across planning methods.

\subsection{Preliminaries}
\label{sec:prelim}
Throughout, we consider a single-phase balanced transmission network modeled as an undirected graph, where~$\mathcal{N}$ denotes the set of buses,~$\mathcal{L}\subset \mathcal{N}\times\mathcal{N}$ the set of transmission branches, and~$\mathcal{G}$ the set of generators, with~$\mathcal{G}_n\subseteq\mathcal{G}$ those at bus~$n$. We denote by~$\mathcal{T}$ a set of representative time periods under study, where period~$t$ is weighted by~$h_t$ hours per year.

In the context of line expansion, we adopt the following notation. Let~$u_{\ell}\in[0,\overline{u}_{\ell}]$ denote the capacity added to branch~$\ell\in\mathcal{L}$, where~$\overline{u}_{\ell}=0$ for branches that are not eligible for expansion, and let the plan vector~$\vu$ collect these values. The feasible plans form the box
\begin{equation}
    \label{eq:feasible_plans}
    \mathcal{U}:=\{\vu:\ \mathbf{0}\le\vu\le\overline{\vu}\}.
\end{equation}
A branch~$\ell$ with existing rating~$\overline{f}_{\ell}$ then has rating~$\overline{f}_{\ell}+u_{\ell}$.

\subsection{Data Center Project Realization Modeling}
Let~$\mathcal{D}$ denote the set of candidate data centers. Each project~$i \in \mathcal{D}$ is characterized by its interconnection bus~$n(i)\in\mathcal{N}$, demand~$d_i \in \mathbb{R}_+$, and externally assigned realization probability~$p_i \in [0,1]$. The realization state of a candidate project~$i$ is represented by a Bernoulli random variable
\begin{equation}
z_i \sim \mathsf{Ber}(p_i),
\end{equation}
where $z_i=1$ indicates that project $i$ is realized and $z_i=0$ indicates that it is not realized.

The project realization probabilities, collected in a vector~$\vp \in [0,1]^{|\mathcal{D}|},$ are treated as planning inputs informed by project maturity or readiness. The present study does not attempt to estimate these probabilities; instead, it examines how transmission expansion decisions change when different probability settings are assumed.

A \emph{realization scenario}~$s$ is a joint outcome~$(z_{is})_{i \in \mathcal{D}}$ of all projects, where~$\boldsymbol{z}_s \in \{0,1\}^{|\mathcal{D}|}$; let~$\mathcal{S}$ denote the set of all such realization scenarios. Under the baseline independence assumption, the probability of scenario~$s\in\mathcal{S}$ is
\begin{equation}
\pi_s(\vp)
=
\prod_{i\in\mathcal{D}}
p_i^{z_{is}}
(1-p_i)^{1-z_{is}},
\label{eq:scenario_probability}
\end{equation}
such that~$\sum_{s \in \mathcal{S}} \pi_s(\vp) = 1$. We write~$\pi_s$ for~$\pi_s(\vp)$ when the argument~$\vp$ is clear from context.

The independence assumption is adopted to isolate the effect of discrete project realization uncertainty. Correlated realizations are left for future work.

The key distinction from conventional continuous load uncertainty is that the uncertain object is the materialization of individual large-load projects. A realization scenario therefore specifies which candidate data center projects materialize and which do not, while preserving their locations and demand magnitudes. This project-realization representation provides the common uncertainty basis used by all planning strategies considered in the paper.

\subsection{From Realization Information to Planning and Evaluation}

Once the data center realization model is established, the same underlying uncertain information can be embedded into different transmission planning strategies. Deterministic planning collapses project uncertainty into a single assumed realization, which depends on how credible planners consider each project to be. Stochastic planning retains multiple realization scenarios and their associated probabilities. Robust planning also considers multiple realization scenarios, but uses the realization set without relying on
probability weighting.

These alternative treatments of the same underlying uncertainty lead to different first-stage expansion plans. These plans are subsequently evaluated under a common ex-post realization environment using the same set of metrics. This separation between planning and evaluation is central to the framework, enabling fair comparison across planning approaches.

\section{Transmission Planning Under Data Center Realization Uncertainty}
\label{sec:planning}

All planning strategies considered in this study are represented using a common two-stage expansion structure. The first stage determines transmission expansion before future data center realizations are known, while the second stage represents scenario-dependent system operation after a realization is observed. The planning strategies differ primarily in how the underlying realization information is incorporated into the first-stage decision. To focus the analysis on the impact of uncertain data center project realization, the present formulation considers capacity expansion of existing transmission corridors, without introducing additional transmission siting decisions. The proposed framework can be extended to include candidate new lines by introducing corresponding first-stage investment decisions and associated network constraints.

\subsection{Unified Two-Stage Transmission Expansion Model}
\label{subsec:unified_tep}

For the planning strategies considered in this paper,  the transmission expansion problem can be written generically as
\begin{equation}
 \min_{\vu\in\mathcal{U}}
\quad
C (\vu)
+
R(\vu)
\label{eq:unified_tep}
\end{equation}
where~$C(\vu)$ is the annualized first-stage transmission investment cost, and~$R(\vu)$ is a second-stage cost that combines recourse costs~$Q_s(\vu)$ of the realization scenarios, both to be defined explicitly below in~\eqref{eq:recourse_problem}. Differences in planning strategies we consider are only in~$R(\cdot)$; see Section~\ref{subsec:planning_strategies}.

A generic investment cost function is
\begin{equation}
C(\vu)
=
\sum_{\ell\in\mathcal{L}}
\left(
c_{\ell}u_{\ell}
+
\kappa_\ell \Ind{u_{\ell} >0}
\right),
\label{eq:investment_cost}
\end{equation}
where~$c_{\ell}$ is the annualized cost per MW of added capacity, and~$\kappa_\ell \ge 0$ is the annualized fixed expansion cost incurred if branch~$\ell$ is expanded at all. 

For a fixed first-stage expansion plan~$\vu \in \mathcal{U},$ the recourse cost~$Q_s(\vu)$ under realization scenario $s$ is the optimal value of a multi-time-period DC optimal power flow with load shedding. We obtain the program

\begin{subequations}
\label{eq:recourse_problem}
\begin{align}
Q_s(\vu)
=
\min_{\substack{\vg_t,\vf_t,\vtheta_t,\vLS_t\\ t\in\mathcal{T}}}\;
&\sum_{t\in\mathcal{T}} h_t
\Bigl[
\sum_{k} C_k(g_{kt})
+
\VOLL\cdot\mathbf{1}^{\top}\vLS_t
\Bigr]
\nonumber
\\
\text{s.t.}\quad
& \mE\vg_t - \mA\vf_t + \vLS_t = \vd^0_t +  \mGamma\vz_s,
\label{eq:power_balance}
\\
& \vf_t = \mB \mA^{\top}\vtheta_t,
\label{eq:dc_power_flow}
\\
& \lvert\vf_t\rvert \le \overline{\vf} + \vu,
\label{eq:thermal_limit}
\\
& \underline{\vg} \le \vg_t \le \overline{\vg},
\label{eq:generation_limit}
\\
& \mathbf{0} \le \vLS_t \le \vd^0_t +  \mGamma\vz_s,
\label{eq:load_shedding_limit}
\end{align}
\end{subequations}
for all~$t \in \mathcal{T}.$ Here, the vector~$\vg_t$ collects the setpoints~$\{g_{kt}\}$ of generators~$k \in \mathcal{G}$, the vector~$\vf_t$ collects the branch flows, the vector~$\vtheta_t$ collects the nodal voltage phase angles, and the vector~$\vLS_t$ collects the load shed at time period~$t \in \mathcal{T}$. The matrix~$\mE \in \{0,1\}^{\vert\mathcal{N}\vert \times \vert \mathcal{G} \vert}$ maps generators to nodes, and the matrix~$\mA\in\{-1,0,1\}^{\vert \mathcal{N}\vert \times \vert \mathcal{L} \vert}$ is the bus-branch incidence matrix of the underlying network. The matrix~$\mB := -\mathsf{diag}(\boldsymbol{b})$ is a diagonal matrix of line series susceptances.

The base system demand is~$\vd^0_t$, and~$\mGamma\in\mathbb{R}_+^{|\mathcal{N}|\times|\mathcal{D}|}$ assigns project demands to their corresponding buses. In particular,
\[
\Gamma_{ni} = \begin{cases}
    d_i & \text{if project~$i$ connects at bus~$n$,}\\
    0 & \text{otherwise}.
\end{cases}
\]
The vectors~$\overline{\vf}$, $\underline{\vg}$ and~$\overline{\vg}$ collect the existing branch thermal and generator limits. In all constraints of the program~\eqref{eq:recourse_problem}, the inequalities and~$\lvert\cdot\rvert$ hold elementwise. Each generator~$k$ has a production cost function~$C_k(\cdot)$, and~$\VOLL$ is a prespecified scalar describing the value of lost load.
The scenario enters the program~\eqref{eq:recourse_problem} only through the term~$\mGamma\vz_s$ in constraints~\eqref{eq:power_balance} and~\eqref{eq:load_shedding_limit}, and the planning variable~$\vu$ through~\eqref{eq:thermal_limit}.

The generic formulation allows production cost to be included in the objective function. In the case studies, however, production cost is intentionally excluded in order to isolate the transmission investment--reliability tradeoff.

\subsection{Alternative Treatments of Realization Uncertainty}
\label{subsec:planning_strategies}

The unified model in \eqref{eq:unified_tep} allows alternative planning strategies to be represented through different choices of the second stage cost~$R(\cdot)$. All methods share the same transmission network, investment representation, and operational model, while differing in their treatment of the data center realization information.

\subsubsection{Deterministic All-Project Planning (Det-All)}

Det-All assumes that every candidate data center project materializes. Let $s_{\mathrm{all}}$ denote the corresponding realization, such that

\begin{equation}
z_{i,s_{\mathrm{all}}}=1,
\qquad
\forall i\in\mathcal{D},
\label{eq:det_all_state}
\end{equation}

and the generic objective becomes

\begin{equation}
 \min_{\vu\in\mathcal{U}}
\quad
C (\vu)
+
Q_{s_{\mathrm{all}}} (\vu).
\label{eq:det_all}
\end{equation}

Det-All therefore represents a conservative deterministic treatment of data center realization uncertainty.

\subsubsection{Deterministic Threshold-screened Planning (Det-$\tau$)}

For a probability threshold $\tau \in (0,1]$, the deterministic realization used for planning is defined as

\begin{equation}
z_i^{(\tau)}
=
\mathbf{1}
\left(
p_i\ge\tau
\right),
\qquad
\forall i\in\mathcal{D}.
\label{eq:threshold_state}
\end{equation}

where candidate projects with realization probability below $\tau$ are excluded from the planning scope.  

Let $s_{\tau}$ denote the resulting scenario. The problem becomes

\begin{equation}
 \min_{\vu\in\mathcal{U}}
\quad
C (\vu)
+
Q_{s_{\tau}} (\vu).
\label{eq:det_threshold}
\end{equation}

This formulation provides a stylized representation of the project screening process used in planning practice based on project readiness or maturity. 

\subsubsection{Stochastic Planning (STO)}

STO retains a set of realization scenarios and their associated probabilities. Thus, the resulting stochastic TEP problem becomes
\begin{equation}
 \min_{\vu\in\mathcal{U}}
\quad
C (\vu)
+
\sum_{s\in\mathcal{S}}
\pi_s Q_s (\vu).
\label{eq:stochastic_tep}
\end{equation}

\subsubsection{Robust Planning}

Robust planning uses the same realization uncertainty set but does not use realization probabilities in the objective. Instead, it protects against the realization producing the highest recourse cost.

The robust planning problem is therefore
\begin{equation}
 \min_{\vu\in\mathcal{U}}
\quad
C (\vu)
+
\max_{s\in\mathcal{S}}
Q_s (\vu).
\label{eq:robust_tep}
\end{equation}
Unlike Det-All, which optimizes for one predefined all-project realization scenario, robust planning, as in~\eqref{eq:robust_tep}, determines the most adverse realization from the uncertainty set.

Table~\ref{tab:planning_strategy_summary} summarizes the different treatments of the common data center realization information.

\begin{table}[t]
\caption{Treatment of Data Center Realization Uncertainty}
\label{tab:planning_strategy_summary}
\centering
\begin{tabular}{p{0.18\columnwidth}
                p{0.4\columnwidth}
                p{0.28\columnwidth}}
\toprule
Method &
Planning realization set &
Second-stage cost~$R$ \\
\midrule

Det-All &
All-project realization &
$Q_{s_{\mathrm{all}}}$ \\

Det-$\tau$ &
Threshold-screened realization &
$Q_{s_{\tau}}$ \\

STO &
Multiple realization scenarios &
$\sum_s \pi_s Q_s$ \\

Robust &
Multiple realization scenarios &
$\max_s Q_s$ \\

\bottomrule
\end{tabular}
\end{table}

\subsection{Solution Implementation}
\label{subsec:solution_implementation}

The stochastic planning problem is solved using a warm-started multicut Benders decomposition framework~\cite{van1969shaped,alizadeh2016efficient}. The first-stage transmission expansion variables are retained in the master problem, while the operational variables are assigned to linear recourse subproblems. Dual information from the recourse subproblems is used to generate Benders optimality cuts. The implementation first solves a relaxed-master phase to obtain a warm-start solution and subsequently solves the mixed-integer master problem until the prescribed convergence tolerance is satisfied. The robust planning model is implemented as a scenario-based robust problem. It uses the same realization scenario set as the stochastic model, and is solved with the same warm-started Benders decomposition structure. The master problem contains the first-stage expansion variables and an auxiliary worst-case recourse variable, which is constrained by scenario-wise Benders cuts generated from the dual solutions of the linear recourse subproblems.

\section{Unified Ex-Post Evaluation}
\label{sec:evaluation}

The planning strategies introduced in Section~\ref{sec:planning} may produce different first-stage transmission expansion plans. To compare these plans consistently, this section establishes a unified ex-post evaluation framework that separates two conceptually distinct dimensions: \textit{what performance is measured} and \textit{under which realization-probability distribution the performance is evaluated}.

Let~$\hat{\vp} \in [0,1]^{|\mathcal{D}|}$ denote the data center realization probability vector used during the planning stage. The resulting first-stage expansion plan for method $m$
is denoted by~$\vu_{m}\left(\hat{\vp}\right)$.

Once the expansion plan is fixed, it is evaluated under an ex-post probability vector~$\vp$. The overall evaluation can therefore be represented generically as

\begin{equation}
\mPhi_m(\hat{\vp},\vp)
:=
\mPhi\!\left(\vu_m(\hat{\vp}),\vp\right).
\label{eq:method_evaluation_mapping}
\end{equation}
where $\mPhi(\cdot)$ denotes the collection of metrics defined below.

\subsection{Ex-Post Performance Metrics}
\label{subsec:performance_metrics}

The same set of performance metrics is applied under both correct-probability and cross-probability evaluation. These metrics characterize the transmission investment, reliability performance, economic consequence, and future adaptation requirement of a fixed expansion plan.

\subsubsection{Investment}

For a fixed expansion plan $\vu_{m}$, the annualized transmission investment is

\begin{equation}
C\left(\vu_{m}\right).
\label{eq:eval_initial_investment}
\end{equation}

\subsubsection{Expected Energy not Served (EENS)}
For realization scenario $s$, the energy not served (ENS) is

\begin{equation}
\ENS_s\left(\vu_{m}\right)
=
\sum_{t\in\mathcal{T}}
h_t\cdot\!\left( 
\mathbf{1}^{\top}\vLS_t
\right)
\label{eq:scenario_ens}
\end{equation}

where the expansion plan is fixed and the system is redispatched under the realized data center scenario.

Given an evaluation probability vector $\vp$, the EENS is

\begin{equation}
\EENS
\left(
\vu_{m},
\vp
\right)
=
\sum_{s\in\mathcal{S}}
\pi_s
\left(
\vp
\right)
\ENS_s\left(\vu_{m}\right),
\label{eq:eval_eens}
\end{equation}

\subsubsection{Expected Total Cost (ETC)}

The expected total cost is defined as

\begin{equation}
\ETC
\left(
\vu_{m},
\vp
\right)
=
C\left(\vu_{m}\right)
+
\VOLL
\cdot
\EENS
\left(
\vu_{m},
\vp
\right).
\label{eq:eval_etc_general}
\end{equation}

When generation production cost is included in the objective function, \eqref{eq:eval_etc_general} is extended to include the expected generation cost under the same evaluation
distribution.

\subsubsection{Subsequent Adaptation (SA)}

To quantify the future corrective expansion requirement after the actual data center realization becomes known, the subsequent adaptation cost for realization scenario $s$ is defined
as

\begin{equation}
\SA_s\left(\vu_{m}\right)
=
\min_{\Delta\vu\ge\mathbf{0}}
\quad
C(\vu_{m}+\Delta\vu)-C(\vu_{m}),
\label{eq:scenario_adaptation}
\end{equation}
subject to

\begin{equation}
\mathbf{0}
\le
\Delta\vu
\le
\overline{\vu}-\vu_m,
\label{eq:adaptation_capacity_limit}
\end{equation}

together with~\eqref{eq:power_balance}--\eqref{eq:load_shedding_limit} for scenario~$s$ with~$\vu_m+\Delta\vu$ in place of~$\vu$ and~$\vLS_t\gets \mathbf{0}$.

The initial expansion plan $\vu_{m}$ is fixed and considered to be implemented. The adaptation problem therefore determines the minimum additional investment required to eliminate load shedding after the realized data center scenario is known.

The expected subsequent adaptation (ESA) under evaluation probability vector $\vp$ is

\begin{equation}
\ESA
\left(
\vu_{m},
\vp
\right)
=
\sum_{s\in\mathcal{S}}
\pi_s
\left(
\vp
\right)
\SA_s\left(\vu_{m}\right).
\label{eq:expected_adaptation}
\end{equation}

Accordingly, the ex-post evaluation vector can be summarized as
\begin{equation}
\mPhi_m(\hat{\vp},\vp)
=
\left[
C,
\EENS,
\ETC,
\ESA
\right].
\label{eq:metric_vector}
\end{equation}

\subsection{Correct-Probability Evaluation}
\label{subsec:correct_probability_evaluation}

The first evaluation regime assumes that the realization probabilities used during planning are correct, i.e.,~$\hat{\vp} = \vp$.

For planning method $m$, correct-probability performance is therefore

\begin{equation}
\mPhi^{\mathrm{corr}}_m(\mathbf{p})
=
\mPhi
(
\vu_{m}(\mathbf{p}),
\mathbf{p}
).
\label{eq:correct_probability_metrics}
\end{equation}

All planning strategies are evaluated under the same realization distribution even though the probability information may have been used differently during the planning stage.

\textit{Remark:} Under correctly specified probabilities, STO minimizes the same ETC metric used in ex-post evaluation. Its ETC optimality under this regime therefore follows directly from the stochastic objective. Importantly, this property does not imply optimality in either investment or EENS. In particular, STO is not required to have either lower investment or lower EENS than other alternative planning strategies. A more conservative plan may achieve lower EENS through higher investment, while a more selective deterministic plan may reduce investment at the cost of larger reliability exposure.

\subsection{Misspecified Model Evaluation}
\label{subsec:cross_probability_evaluation}

The second evaluation regime relaxes the assumption that the realization probabilities used for planning are correct. The plan is built under~$\hat{\vp}$ but evaluated under the true probability vector~$\vp \neq \hat{\vp}$. Its cross-probability performance is
\begin{equation}
\mPhi_m(\hat{\vp},\vp),\quad \mathrm{where}\quad \hat{\vp} \neq \vp.
\label{eq:cross_probability_metrics}
\end{equation}

Unlike the well-specified case, stochastic planning is not guaranteed to minimize ETC when it is evaluated under a different probability distribution $\vp$. Cross-probability evaluation therefore quantifies how sensitive the performance of each planning strategy is to probability errors.

To distinguish mild probability errors from substantial misspecification, the distance between the planning and true probability vectors is measured by

\begin{equation}
\delta
=
\left\|
\hat{\vp}
-
\vp
\right\|_{\infty}
=
\max_{i\in\mathcal{D}}
\left|
\hat{p}_i
-
p_i
\right|,
\label{eq:probability_distance}
\end{equation}

The following numerical study groups cross-probability cases into Small, Moderate and Severe subgroups according to the magnitude of $\delta$. 

\begin{equation}
\begin{aligned}
\text{Small error:}\quad
& \delta \le 0.2,\\
\text{Moderate error:}\quad
& 0.2 < \delta \le 0.5,\\
\text{Severe error:}\quad
& \delta > 0.5.
\end{aligned}
\label{eq:severity_groups}
\end{equation}

\section{Numerical Studies}
\label{sec:numerical}

This section evaluates the proposed framework on two transmission test systems. A synthetic 21-bus Mid-Atlantic system \cite{Shao2025PJM21} is used as the primary test system for detailed analysis of planning behavior, correct-probability, cross-probability, and subsequent adaptation performance. A modified IEEE 24-bus system \cite{IEEE1999RTS} is then used to examine whether the principal planning insights persist under a different network topology.

\subsection{Test Systems and Experimental Setup}
\label{subsec:test_setup}

\begin{figure}[t]
\centering
\includegraphics[width=0.8\linewidth]{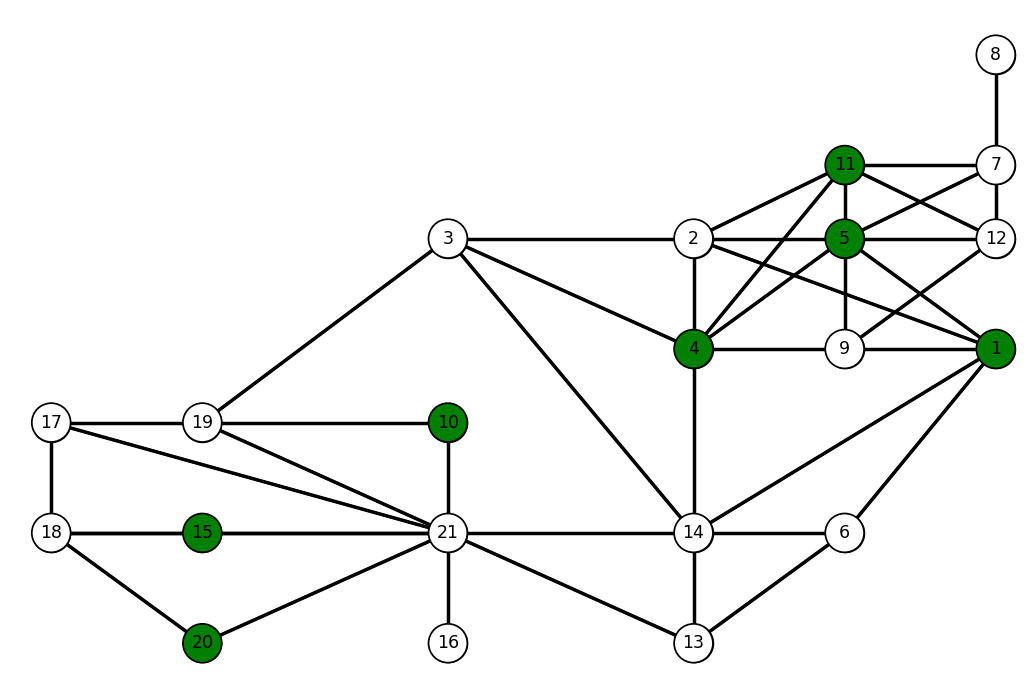}
\caption{Synthetic 21-bus Mid-Atlantic system topology. Nodes marked in green represent those with data center interconnection requests.}
\label{fig:pjm21_topology}
\end{figure}

The primary test system is a synthetic 21-bus system, representing a large interconnected transmission network centered in Mid-Atlantic region of eastern United States as shown in Fig.~\ref{fig:pjm21_topology}. Each bus represents an aggregated subregion. The resulting system provides substantial geographic diversity and interregional transfer constraints while remaining computationally tractable.

Seven buses are selected to be candidate data center locations as marked in green in Fig.~\ref{fig:pjm21_topology}. To evaluate planning performance over a broad probability space, the realization probabilities of all seven candidate data centers are allowed to vary over the set $\{0.1,0.3,0.5,0.8\}$. A total of 397 probability vectors are generated through a balanced discrete randomized sampling procedure to obtain broad coverage of the probability space while maintaining computational tractability. The same set of probability vectors is used for all planning methods. 

All reported performance metrics are evaluated on an annual basis. Monetary quantities, including investment, expected total cost, and subsequent adaptation cost, are expressed in M\$, while ENS is expressed in MWh, unless otherwise specified.

\begin{table}[t]
\caption{Main Settings of the 21-Bus Case Study}
\label{tab:pjm21_setup}
\centering
\renewcommand{\arraystretch}{1.10}

\begin{tabular}{
>{\raggedright\arraybackslash}m{0.42\columnwidth}
>{\raggedright\arraybackslash}m{0.50\columnwidth}
}
\toprule
Parameter & Setting \\
\midrule

Candidate data center buses
& 1, 4, 5, 10, 11, 15, 20 \\

Data center size
& 3000 MW/site \\

Base probability vector
& $(0.1,0.8,0.8,0.1,0.8,0.8,0.8)$ \\

Possible probability values
& $\{0.1,0.3,0.5,0.8\}$ \\

Number of probability settings
& 397 unique vectors \\

Base VOLL
& \$10,000/MWh \\

Variable expansion cost
& \$21,940/(MW$\cdot$km) \\

Fixed expansion cost
& \$5 million \\

Representative periods
& 4 high-load and 4 quantile blocks, weighted by $h_t$ \\

\bottomrule
\end{tabular}

\end{table}

For each probability setting, the planning strategies considered in Section~\ref{sec:planning} are used to obtain first-stage expansion plans. The resulting expansion plans are then fixed and evaluated using the common ex-post framework as in Section~\ref{sec:evaluation}. The primary comparisons focus on STO, Det-All, and Det-$\tau$ with $\tau\in\{0.3,0.5,0.8\}$. Under the 21-bus system settings, Robust and Det-All produce the same expansion plan and are therefore reported together in the
corresponding results.

\subsection{Performance Under Correctly Specified Probabilities}
\label{subsec:correct_results}

The first set of experiments assumes that the planning probabilities are identical to the true probabilities.

\subsubsection{Illustrative Deterministic Planning Dilemma}
\label{subsubsec:illustrative}

Consider the base probability setting [0.1, 0.8, 0.8, 0.1, 0.8, 0.8, 0.8], two representative deterministic planning approaches reflecting common planning practices are:

\begin{itemize}
    \item Det-All: assumes all candidate data centers materialize;
    \item Det-$\tau$: neglects the low-probability data centers at buses 1 and 10, retaining only the other five data centers.
\end{itemize}

The uncertain data center scenarios collapse into single fixed realization. By contrast, stochastic planning keeps all possible data centers with their respective probability information.

Results are shown in Table~\ref{tab:illustrative_dilemma}, including required investment, and ENS under several representative scenarios:

\begin{itemize}
    \item $s_0$: only the five high-probability data center projects
    at buses 4, 5, 11, 15, and 20 materialize;

    \item $s_{10}$: the data center project at bus 10
    additionally materializes on top of $s_0$;

    \item $s_{1}$: the data center project at bus 1
    additionally materializes on top of $s_0$;

    \item $s_{\mathrm{all}}$: all seven candidate data center projects materialize.
\end{itemize}

\begin{table}[t]
\caption{Illustrative Example Under Alternative Data Center Realizations}
\label{tab:illustrative_dilemma}
\centering
\renewcommand{\arraystretch}{1.10}
\setlength{\tabcolsep}{3pt}

\begin{tabular}{
>{\raggedright\arraybackslash}m{0.22\columnwidth}
>{\centering\arraybackslash}m{0.16\columnwidth}
cccc
}
\toprule

\multirow{2}{*}{Planning Strategy}
&
\multirow{2}{*}{\shortstack{Investment\\(M\$)}}
&
\multicolumn{4}{c}{Energy Not Served (MWh)}
\\

\cmidrule(lr){3-6}

&
&
$s_0$
&
$s_{10}$
&
$s_1$
&
$s_{\mathrm{all}}$
\\

\midrule

Det-All
& 43.55
& 0
& 0
& 0
& 0
\\

Det-$\tau$
& 3.33
& 0
& 8,991
& 121,198
& 132,570
\\

STO
& 18.52
& 0
& 2,685
& 5,555
& 12,956
\\

\bottomrule
\end{tabular}
\end{table}

The two deterministic ways represent contrasting treatments to the same realization uncertainty. Det-All commits \$43.55 million investment and eliminates ENS across all representative realizations, whereas Det-$\tau$ reduces investment to \$3.33 million but becomes highly risk-exposed when either excluded data center subsequently materializes.

STO provides an intermediate solution to this dilemma. With \$18.52 million of investment, it requires substantially less investment commitment than Det-All while greatly reducing ENS compared to Det-$\tau$.

\subsubsection{Aggregate Economic and Reliability Performance}
\label{subsubsec:correct_aggregate}

The illustrative example is next extended to a broader range of probability settings. Table~\ref{tab:correct_probability} summarizes the mean economic and reliability metrics under correct probabilities.

\begin{table}[t]
\caption{Performance Under Correctly Specified Probabilities}
\label{tab:correct_probability}
\centering
\renewcommand{\arraystretch}{1.10}
\setlength{\tabcolsep}{4pt}

\begin{tabular}{
>{\raggedright\arraybackslash}m{0.30\columnwidth}
ccccc
}
\toprule
Metric
& STO
& \makecell{Det-All/\\Robust}
& Det-0.8
& Det-0.5
& Det-0.3
\\
\midrule
Mean investment (M\$)
& 26.37
& 43.55
& 7.56
& 16.49
& 28.30
\\

Mean ETC (M\$)
& 31.12
& 43.55
& 499.14
& 219.66
& 72.72
\\

Mean EENS (MWh)
& 474
& 0
& 49,158
& 20,317
& 4,442
\\

Mean ESA (M\$)
& 1.38
& 0.00
& 7.89
& 3.73
& 0.89
\\

Mean expansion (MW)
& 1,090
& 1,364
& 348
& 677
& 990
\\

\bottomrule
\end{tabular}
\end{table}

\begin{figure}[t]
    \centering
    \includegraphics[width=0.75\linewidth]{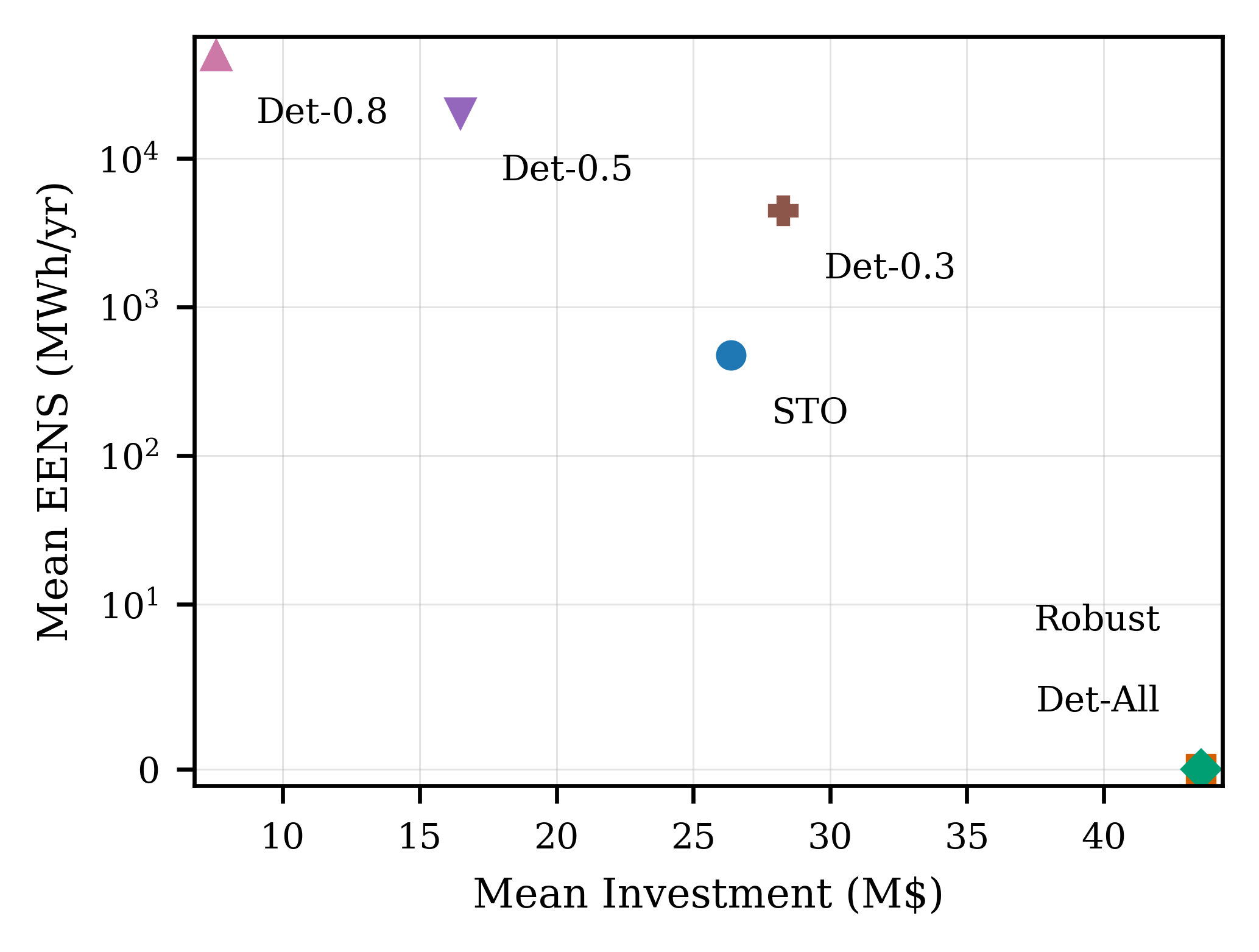}
    \caption{Mean investment versus mean EENS across the correctly specified probability settings. }
    \label{fig:correct_scatter}
\end{figure}

The results reveal a general investment--reliability tradeoff across the planning strategies as in Fig.~\ref{fig:correct_scatter}: larger investment is associated with lower load-shedding exposure. For Det-$\tau$ strategies, a higher screening threshold excludes more candidate projects from the planning, resulting in lower investment but higher EENS, whereas reducing the threshold includes more projects, increases investment, and reduces EENS.

STO requires substantially less investment than Det-All, while avoiding the much larger EENS of Det-$\tau$ plans. As expected, STO achieves the lowest mean ETC across the tested planning strategies.

\subsubsection{Subsequent Adaptation}
\label{subsubsec:correct_adaptation}

The subsequent-adaptation metric provides an additional perspective on the planning tradeoff. As shown in Table~\ref{tab:correct_probability}, Det-All makes the largest investment and requires no subsequent adaptation in the tested cases. In contrast, the more selective Det-0.8 and Det-0.5 strategies reduce investment but incur substantially larger expected corrective expansion later.

These results show that plans with low investment can shift part of the expansion burden into future corrective expansion. STO again provides an intermediate initial-versus-future investment planning position: it commits more transmission investment than aggressive Det-0.8 and Det-0.5 strategies but substantially reduces the corresponding adaptation requirement. Det-0.3, by contrast, makes slightly more initial investment than STO and correspondingly achieves lower ESA.

\subsection{Performance Under Probability Misspecification}
\label{subsec:misspecified_results}
The preceding analysis assumes that the realization probabilities used in planning are correctly specified. In practice, however, planning probabilities may be based on incomplete project information and may differ from the true realization probabilities.

Let $\mathcal{P}=\{\mathbf{p}^{(1)},\ldots,\mathbf{p}^{(397)}\}$ denote the sampled probability settings. For each ordered pair $(a,b)$ with $a\neq b$, an expansion plan constructed using $\mathbf{p}^{(a)}$ as the planning probability vector is evaluated under $\mathbf{p}^{(b)}$ as the true probability vector. This produces $157{,}212$ ordered cross-probability cases. All cases are weighted equally when reporting aggregate results. 

\subsubsection{Aggregate Cross-Probability Performance}
\label{subsubsec:cross_aggregate}

Table~\ref{tab:misspecified} summarizes the mean results over all cross-probability cases. For a given metric, the STO win fraction is defined as the fraction of planning/true probability pairs for which STO achieves a strictly lower value than the specified baseline. Ties are not counted as wins.

\begin{table}[t]
\caption{Aggregate Performance Under Probability Misspecification}
\label{tab:misspecified}
\centering
\resizebox{\columnwidth}{!}{%
\begin{tabular}{lrrrrr}
\toprule
Metric &
STO &
\makecell{Det-All/\\Robust} &
Det-0.8 &
Det-0.5 &
Det-0.3 \\
\midrule

Mean investment (M\$)
& 26.37 & 43.55 & 7.56 & 16.49 & 28.30 \\

Mean ETC (M\$)
& 37.21 & 43.55 & 701.39 & 446.62 & 217.48 \\

STO win fraction, ETC
& -- & 82.47\% & 98.97\% & 96.15\% & 91.93\% \\

Mean EENS (MWh)
& 1,084 & 0 & 69,384 & 43,013 & 18,918 \\

STO win fraction, EENS
& -- & 0.00\% & 99.50\% & 92.92\% & 59.28\% \\

Mean ESA (M\$)
& 2.62 & 0.00 & 10.63 & 7.32 & 3.64 \\

STO win fraction, ESA
& -- & 0.00\% & 99.50\% & 91.62\% & 50.85\% \\

\bottomrule
\end{tabular}%
}
\end{table}

Probability misspecification materially affects the strategies whose expansion decisions depend on the assumed realization probabilities, including STO and Det-$\tau$. Det-All and Robust, by contrast, are independent of the probability vector. Over the cross-probability cases, STO retains a lower mean ETC than Det-All/Robust and all three Det-$\tau$ strategies. Its mean EENS remains above the zero-EENS Det-All/Robust benchmark but is substantially below those of the Det-$\tau$ strategies. 

The pairwise results provide a complementary view. STO achieves a lower ETC than Det-0.8, Det-0.5, and Det-0.3 in 98.97\%, 96.15\%, and 91.93\% of the tested cross-probability cases, respectively. The corresponding EENS win fractions are 99.50\%, 92.92\%, and 59.28\%, indicating a particularly strong advantage over the more selective Det-$\tau$ strategies.

\subsubsection{Misspecification Severity Analysis}
\label{subsubsec:severity}
\begin{figure}
    \centering
    \includegraphics[width=0.7\linewidth]{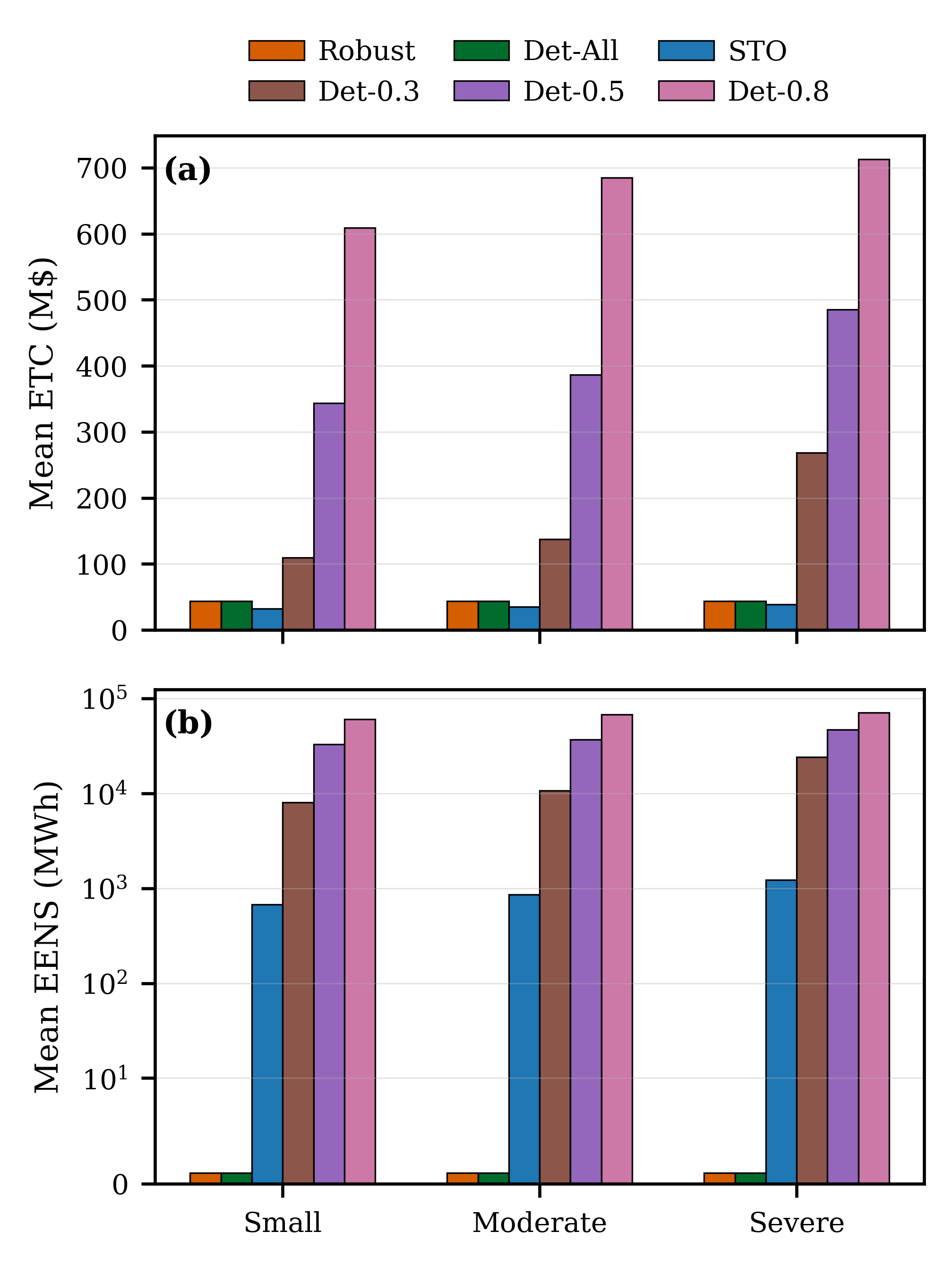}
    \caption{Economic and reliability performance under different probability-misspecification severity levels. Note: For visualization purposes, zero-EENS bars for Robust and Det-All in panel (b) are shown with a small visual height.}
    \label{fig:misspec_severity}
\end{figure}

To further analyze the performance of the planning strategies under different degrees of probability misspecification, the cross-probability cases are grouped according to the probability-distance defined in Section~\ref{sec:evaluation}. The small, moderate, and severe groups contain
1,148, 59,960, and 96,104 probability pairs, respectively. The corresponding mean ETC and EENS values are reported in Table~\ref{tab:severity_etc} and \ref{tab:severity_eens}. Fig.~\ref{fig:misspec_severity} visualizes the economic and reliability performance within each misspecification severity group. 

Within each misspecification-severity group, STO maintains a lower mean ETC than Det-All/Robust and all three Det-$\tau$ strategies. In terms of reliability, STO has higher mean EENS than the most conservative Det-All/Robust benchmark in the 21-bus system, but consistently lower mean EENS than Det-0.8, Det-0.5, and Det-0.3 within each group. These results show that the relative economic and reliability advantages of STO over the deterministic threshold-screening strategies persist across the examined misspecification groups, indicating that STO still provides a more favorable investment--reliability tradeoff than Det-$\tau$ across the examined probability cases.

\begin{table}[t]
\caption{Mean ETC (M\$) Versus Probability-Misspecification Severity}
\label{tab:severity_etc}
\centering
\begin{tabular}{lrrrrr}
\toprule
Severity &
STO &
\makecell{Det-All/\\Robust} &
Det-0.8 &
Det-0.5 &
Det-0.3 \\
\midrule
Small
& 32.08 & 43.55 & 608.88 & 343.45 & 109.61 \\

Moderate
& 35.27 & 43.55 & 684.72 & 386.52 & 137.57 \\

Severe
& 38.48 & 43.55 & 712.90 & 485.35 & 268.62 \\
\bottomrule
\end{tabular}
\end{table}

\begin{table}[t]
\caption{Mean EENS (MWh) Versus Probability-Misspecification Severity}
\label{tab:severity_eens}
\centering
\begin{tabular}{lrrrrr}
\toprule
Severity &
STO &
\makecell{Det-All/\\Robust} &
Det-0.8 &
Det-0.5 &
Det-0.3 \\
\midrule
Small
& 679 & 0 & 60,473 & 33,035 & 8,018 \\

Moderate
& 855 & 0 & 67,789 & 36,937 & 10,686 \\

Severe
& 1,231 & 0 & 70,485 & 46,923 & 24,183 \\
\bottomrule
\end{tabular}
\end{table}

\subsubsection{Subsequent Adaptation Under Misspecified Probabilities}
\label{subsubsec:cross_adaptation}
Probability misspecification increases mean ESA for both STO and the Det-\(\tau\) strategies compared to the correct-probability case. The increase is smaller for STO than for Det-0.8, Det-0.5, and Det-0.3 plans. Under misspecified probabilities, STO also has the lowest mean ESA among these probability-dependent strategies. Its pairwise ESA advantage is particularly strong compared to Det-0.8 and Det-0.5 plans, with win fractions of 99.50\% and 91.62\%, whereas its advantage over the more conservative Det-0.3 plan is less pronounced, with a win fraction of 50.85\%.

\subsection{Model Sensitivities to VOLL}
\label{subsec:sensitivity}

The value of lost load (VOLL) \cite{ovaere2019detailed} affects the STO solution by controlling the penalty assigned to unserved energy in the probability-weighted recourse objective, thereby influencing the tradeoff between investment and reliability. A higher $\VOLL$ increases the incentive to make additional investment when such investment can reduce expected load shedding. 

\begin{figure}
    \centering
    \includegraphics[width=0.75\linewidth]{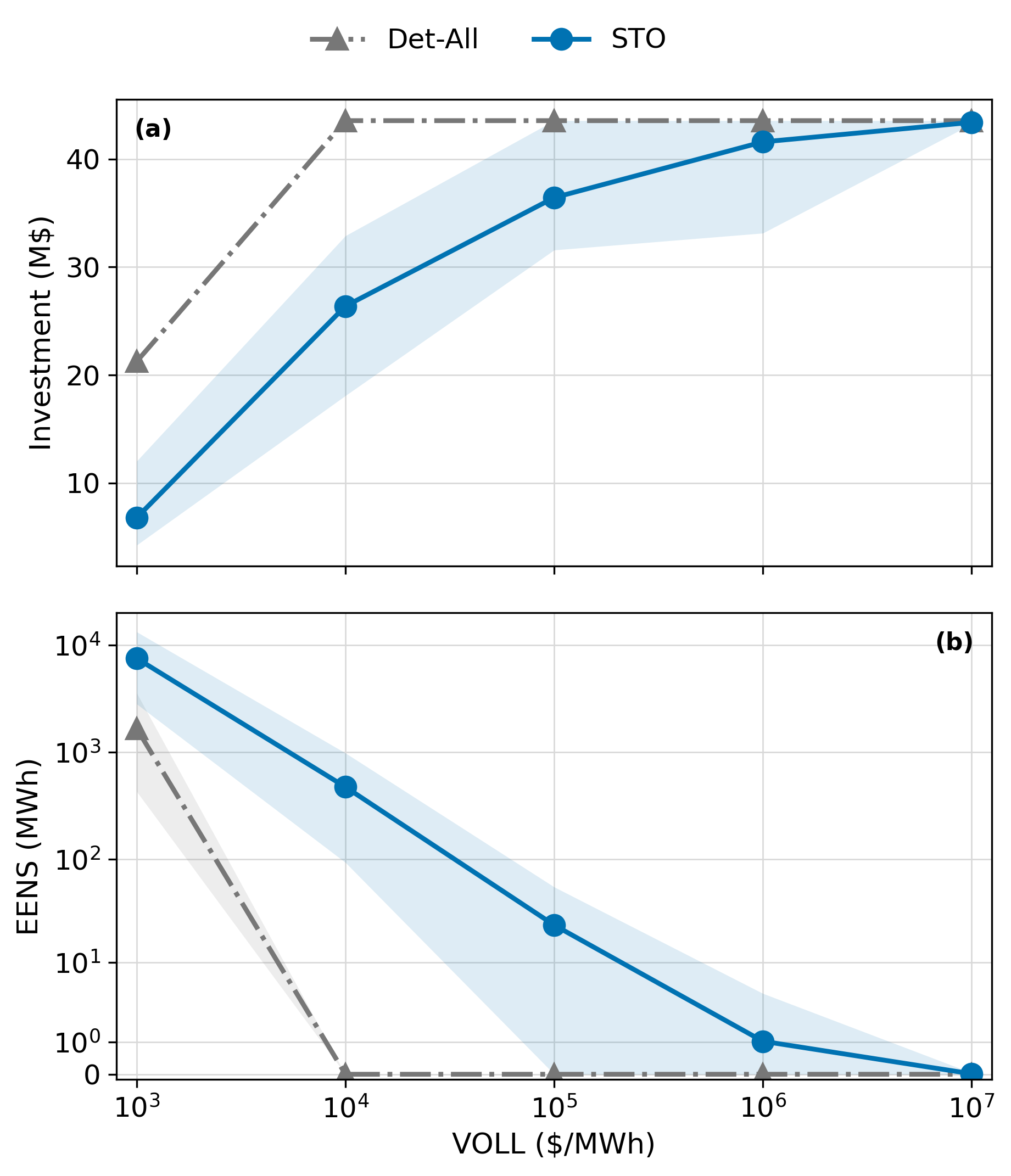}
    \caption{Sensitivity of (a) investment and (b) EENS to VOLL in the 21-bus system. Lines show the means across the examined probability settings, and shaded regions indicate the 5th--95th percentile ranges.}
    \label{fig:voll_sensitivity}
\end{figure}

To examine this effect, the VOLL sensitivity analysis is conducted across the studied probability settings. Fig.~\ref{fig:voll_sensitivity} presents the mean investment and EENS, together with their 5th--95th percentile ranges at different VOLL levels. 

As VOLL increases, STO places greater weight on reducing load shedding across the realization scenarios. Consequently, the optimal STO solution shifts toward higher transmission investment and lower expected energy not served. This pattern reflects the investment--reliability tradeoff embedded in the stochastic objective: at relatively low VOLL, the model may accept limited load shedding in low-probability or high-cost realizations, whereas a higher VOLL makes additional transmission reinforcement increasingly attractive for reducing such reliability exposure.

At sufficiently high VOLL, the STO solution becomes increasingly reliability dominated. In the 21-bus system test case, the numerical results show that the investment and EENS of STO approach those of the conservative Det-All benchmark as VOLL becomes larger.

\subsection{Cross-System Validation on the IEEE 24-bus System}
\label{subsec:ieee24_validation}

A modified IEEE 24-bus system is used as an additional case study to examine whether the planning behavior observed on the 21-bus system persists under a different topology. Based on the original system data, all generator capacities are uniformly scaled by a factor of 2.8, while the base system loads are uniformly scaled by a factor of 2.3 to create transmission reinforcement needs while maintaining sufficient aggregate generation capacity. A zero-$P_{\min}$ simplification is also adopted for all generators. Similarly to the 21-bus system case study, seven candidate data center projects are introduced at buses 1, 4, 8, 10, 14, 19, and 20, each with a capacity of 200 MW. The same 397 probability settings are considered in the planning experiments. Unless otherwise specified, the probability sampling procedure, VOLL, expansion cost assumptions, representative-period construction, and expansion capacity limits are the same as those used in the 21-bus case.

Table~\ref{tab:ieee24_correct} summarizes the aggregate results when the planning and evaluation probabilities are correctly specified.

\begin{table}[t]
\caption{IEEE 24-Bus Validation Results Under Correct Probabilities}
\label{tab:ieee24_correct}
\centering
\resizebox{\columnwidth}{!}{%
\begin{tabular}{lrrrrrr}
\toprule
Metric &
STO &
Robust &
Det-All &
Det-0.8 &
Det-0.5 &
Det-0.3 \\
\midrule

Mean ETC (M\$)
& 52.38 & 77.88 & 80.99 & 380.05 & 143.03 & 73.97 \\

Mean investment (M\$)
& 43.51 & 76.76 & 74.07 & 13.40 & 31.74 & 52.44 \\

Mean EENS (MWh)
& 886 & 113 & 692 & 36,665 & 11,129 & 2,153 \\

Mean expansion (MW)
& 308 & 512 & 493 & 91 & 216 & 357 \\

\bottomrule
\end{tabular}%
}
\end{table}

\begin{figure}[t]
    \centering
    \includegraphics[width=0.75\linewidth]{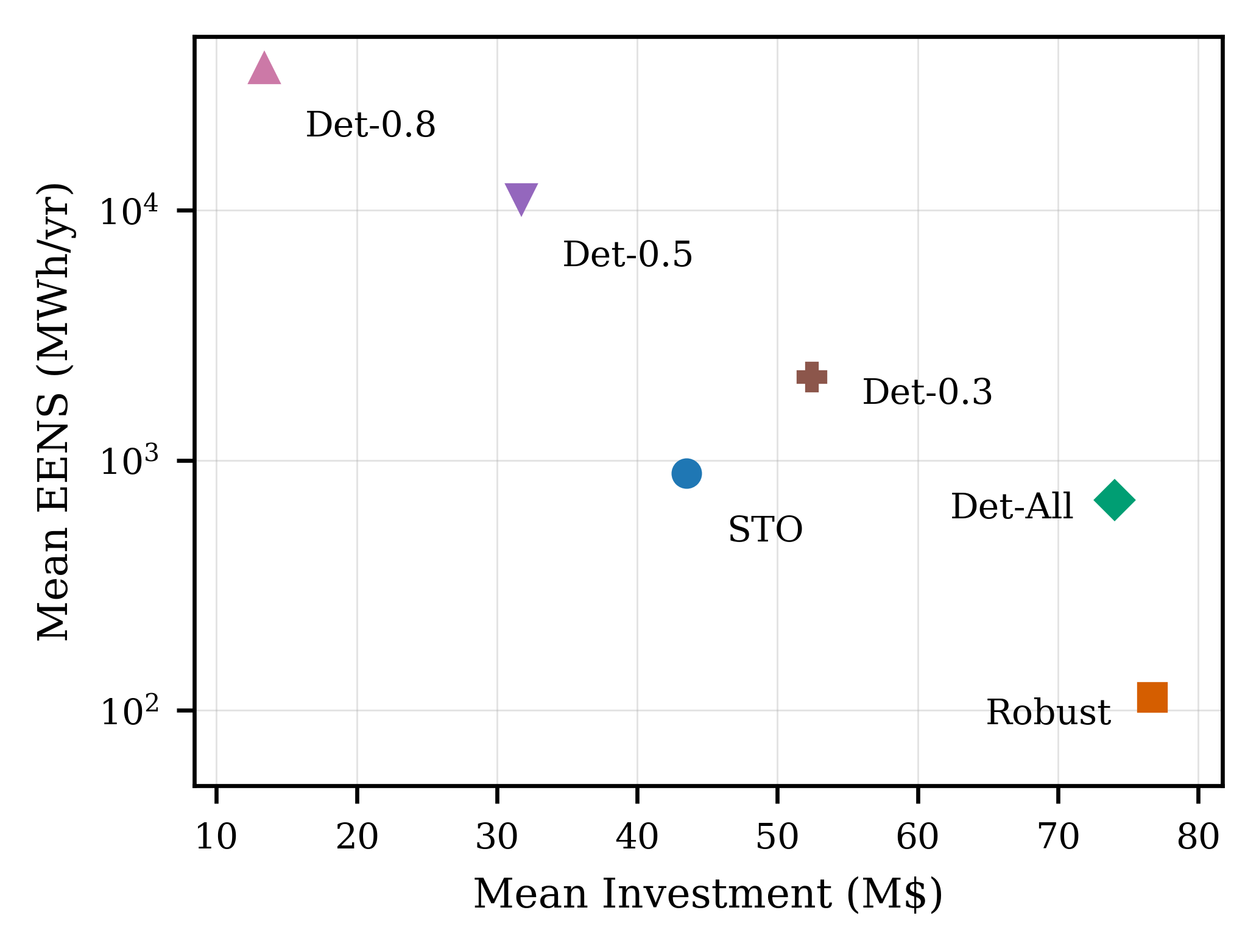}
    \caption{Investment--EENS tradeoff on the modified IEEE 24-bus system.}
    \label{fig:IEEE24-scatter}
\end{figure}

The IEEE 24-bus results reproduce the main investment--reliability pattern observed in the 21-bus system. As illustrated in Fig.~\ref{fig:IEEE24-scatter}, STO again occupies a more balanced intermediate position between the conservative strategy and more selective Det-$\tau$ strategies, and achieves the lowest mean ETC under correct probabilities. Unlike the 21-bus case, the results also show that Robust and Det-All produce different expansion plans which can arise from network effects such as counterflow. Scenario-level inspection shows that the worst-case realization for the robust plan is not the all-project realization scenario. This indicates that the recourse cost is not necessarily monotonic to the number of realized projects.

To further examine cross-system consistency, the resulting plans are also evaluated under probability misspecification. The cross-probability results in Table~\ref{tab:ieee24_misspec} exhibit patterns consistent with those observed in the 21-bus system. Probability misspecification increases both mean ETC and mean EENS for the probability-dependent strategies, including STO and Det-$\tau$. Nevertheless, STO continues to exhibit the lowest mean ETC and substantially lower mean EENS than Det-$\tau$ strategies under probability errors.

\begin{table}[t]
\caption{IEEE 24-Bus Performance Under Probability Misspecification}
\label{tab:ieee24_misspec}
\centering
\resizebox{\columnwidth}{!}{%
\begin{tabular}{lrrrrrr}
\toprule
Metric &
STO &
Robust &
Det-All &
Det-0.8 &
Det-0.5 &
Det-0.3 \\
\midrule

Mean ETC (M\$)
& 70.10 & 77.88 & 80.99 & 567.30 & 298.33 & 142.46 \\

STO win fraction, ETC
& -- & 82.75\% & 84.66\% & 96.71\% & 94.15\% & 90.61\% \\

Mean EENS (MWh)
& 2,659 & 113 & 692 & 55,390 & 26,659 & 9,002 \\

STO win fraction, EENS
& -- & 0.72\% & 27.29\% & 99.82\% & 91.33\% & 58.29\% \\

\bottomrule
\end{tabular}%
}
\end{table}

\section{Discussion and Limitations}
\label{sec:discussion}

The proposed framework is intended to evaluate transmission planning decisions under uncertain data center realization. By retaining project-level realization states and associated probabilities, it enables planners to assess the investment and reliability implications of different planning strategies under consistent evaluation conditions. However, several modeling assumptions should be recognized: realization probabilities are treated as exogenous inputs; the primary experiments exclude generation production cost and consider normal operating conditions without explicit contingency constraints. These assumptions are adopted to keep the analysis focused on project-realization uncertainty; the framework can be extended to calibrated or correlated realization models, and security-constrained operation.

Moreover, the numerical studies are conducted on reduced test systems, leaving scalability to larger networks with larger scenario sets for future work. In addition, the VOLL-based formulation represents load shedding exposure through an economic penalty rather than imposing hard reliability requirements for each realization scenario. Therefore, the framework should be viewed as a decision support approach rather than a replacement for established reliability criteria.

\section{Conclusion}
\label{sec:conclusion}

This paper developed a project-realization-based framework for transmission expansion planning under large, discrete data center realization uncertainty. Candidate projects are represented through fixed locations, requested demand sizes, binary realization states, and assumed realization probabilities. Using this common uncertainty representation, deterministic, stochastic, and robust planning strategies are evaluated within a unified ex-post framework under both correctly specified and misspecified probabilities.

Case studies on two transmission test systems demonstrate the usefulness of the proposed framework for quantitatively assessing and comparing alternative planning strategies. The results reveal a clear overbuilding--underbuilding problem associated with deterministic treatment of uncertain realizations. Stochastic planning uses realization probabilities to hedge across multiple plausible futures, providing a more balanced solution: it requires less investment than the most conservative planning approach while reducing reliability exposure and subsequent adaptation needs relative to the more aggressive high-threshold screening strategies. Under probability misspecification, stochastic planning retains favorable economic and reliability performance relative to the
deterministic threshold-screening strategies across a broad range of misspecified probability cases, continuing to provide a favorable investment--reliability tradeoff.

Future work will focus on empirically grounded realization-probability assessment, correlated project realizations, explicit reliability requirements, contingency-aware planning, and ambiguity-aware or distributionally robust extensions.

\section*{Disclaimer}
The views expressed in this paper are those of the authors and do not reflect the views of PJM Interconnection, L.L.C. or its Board of Managers, of which Le Xie is a member.

\bibliographystyle{IEEEtran}
\bibliography{references}

\end{document}